\documentstyle[prl,aps,twocolumn,epsfig]{revtex}

\draft

\def\be{\begin{equation}}
\def\ee{\end{equation}}
\def\bea{\begin{eqnarray}}
\def\eea{\end{eqnarray}}

\begin{document}

\title{Efficient tight-binding Monte Carlo structural sampling of
complex materials}

\author{Parthapratim Biswas}
\address{Debye Institute,
          Utrecht University, Princetonplein 5, 3508 TA Utrecht,
          the Netherlands}

\author{G.T. Barkema}
\address{Theoretical Physics,
          Utrecht University, Princetonplein 5, 3584 CC Utrecht,
          the Netherlands}

\author{Normand Mousseau}
\address{Department of Physics and Astronomy and CMSS,
          Ohio University, Athens, OH 45701, USA}

\author{W.F. van der Weg}
\address{Debye Institute,
          Utrecht University, Princetonplein 5, 3508 TA Utrecht,
          the Netherlands}

\date{\today}

\maketitle

\begin{abstract}
While recent work towards the development of tight-binding and ab-initio
algorithms has focused on molecular dynamics, Monte Carlo methods
can often lead to better results with relatively little effort. We
present here a multi-step Monte Carlo algorithm that makes use of the
possibility of quickly evaluating local energies. For the thermalization
of a 1000-atom configuration of {\it a}-Si, this algorithm gains
about an order of magnitude in speed over standard molecular dynamics.
The algorithm can easily be ported to a wide range of materials and can
be dynamically optimized for a maximum efficiency.
\end{abstract}
\vskip 0.25cm

Monte Carlo techniques are extensively used in statistical physics,
but have received relatively little attention in the materials theory
community. Most of the recent efforts towards the development of fast
tight-binding or ab-initio codes, for example, has been done within
the framework of molecular dynamics (MD)~\cite{kim,colombo,teter}.
Since all atoms move simultaneously within MD, there are limitations
to the approaches that can be used. In particular, the most promising
order-N MD-based methods~\cite{ordejon,vanderbilt,goedecker} are only
applicable to materials with a well-defined electronic gap, reducing
significantly the number of problems that can be studied with such
algorithms. Monte Carlo (MC) techniques, on the other hand, allow a much
wider range of steps to be used and provide an alternative to MD for a
number of problems.

In this Letter, we introduce such a Monte Carlo algorithm. For a
1000-atom tight-binding model, we find a gain in efficiency of an order
of magnitude over MD without the need for a perfectly clean electronic
gap or any other special consideration.  We first describe the core
of the algorithm and then present some of the details associated with
computing the intermediate steps. Examples are given using a series of
amorphous silicon models described by the tight-binding interaction
of Kwon {\it et al.}~\cite{kwon}.

The general problem that we address here is the generation of a set
of configurations, defined by the atomic positions
$\vec{X}\equiv \{\vec{x}_1, \dots, \vec{x}_N\}$, sampled according to
the microcanonical ensemble, i.e., with a probability proportional to
their Boltzmann weight:
\begin{equation} 
  	P(\vec{X}) \sim \exp(-\beta E(\vec{X})), 
\end{equation}
where $E(\vec{X})$ is the total potential energy of configuration
$\vec{X}$ and $\beta=(k_bT)^{-1}$, the inverse of Boltzmann's constant
$k_b$ times the temperature $T$.  This sampling can be done using standard
MD or MC.

In Monte Carlo, a markovian chain of configurations is generated through a
sequence of trial moves or elementary steps.  A standard move is obtained by
randomly displacing a configuration $\vec{X}_i$ to a trial position
$\vec{X}'_{i+1}=\vec{X}_i+\delta \vec{X}$, shifting one or all atoms in the
box. The trial move is then accepted with a probability given by the Metropolis
criterion~\cite{metrop}
\begin{equation}
P_a={\rm Min} [1, \exp(-\beta \; \delta E)].
\label{eq:metrop}
\end{equation}
where $\delta E=E(\vec{X}'_{i+1})-E(\vec{X}_i)$ is the energy
difference between the trial and the initial positions.

A straightforward implementation of this standard Monte Carlo algorithm
requires thus one full energy calculation per trial step, making it
usually much slower than a standard MD simulation.  It is possible,
however, to speed up the MC simulation significantly, especially as the
cost of computing the total energy increases faster than linear with
the system size.  This improvement exploits the fact that the energy
difference between two configurations, $\delta E$, which differ only
locally, can be estimated in a much quicker way than the total energy,
as discussed below. Using this approximate value, the acceptance ratio,
Eq. \ref{eq:metrop}, is modified as follows:
\begin{eqnarray}
P_a      &=&{\rm Min} \left[1, \exp\left(-\tilde{\beta} \; \delta \tilde{E} 
\right)\right] \nonumber \\
     &\times& {\rm Min} \left[1,
  \exp\left(- \beta \delta E+\tilde{\beta} \delta \tilde{E} \right)\right],
\label{eq:newmetrop}
\end{eqnarray}
where $\tilde{\beta}$ is an inverse temperature close to $\beta$.
Note that detailed balance is still strictly obeyed, irrespectively of the
quality of the estimate and the value of $\tilde{\beta}$.  This expression
can be implemented by first using an accept--reject procedure based on
the first factor, followed by an accept--reject procedure based on the
second factor {\it only} if the first one is accepted. The gain in doing
this is that the first factor in this equation requires only an {\it
estimate} of the energy difference, which is easily computed, while
the second factor, that involves an expensive total energy calculation,
is computed {\it only} if the first one is accepted.  If the estimate
is accurate, $\tilde{\beta}$ can be chosen equal to $\beta$ and the
acceptance probability for the second factor is close to unity; 
effectively a full energy calculation is only required per {\it accepted}
step rather than per {\it trial} step. Since rejected moves are now cheaper,
the trial step size can be enlarged for an increased efficiency.

A further reduction in the number of full energy calculations is possible
by making a sequence of $M$ intermediate displacement steps, each one
accepted or rejected based on the estimate of the energy difference,
followed by a single accept--reject procedure which requires a full
energy calculation.  The resulting acceptance ratio, which still strictly
preserves detailed balance, becomes
\begin{eqnarray}
P_a    &=& \prod_{i=1}^M \left[
   {\rm Min} \left[1, \exp\left(-\tilde{\beta} \delta
\tilde{E}_i\right)\right]
   \right] \nonumber \\
  &\times&
   {\rm Min} \left[1, \exp\left(-\beta \delta E-\sum_{i=1}^M 
\tilde{\beta} \delta \tilde{E} \right)\right].
\label{eq:multimetrop}
\end{eqnarray}
These two variables,
the number and typical size of the intermediate trial moves $M$, can then
be optimized for speed.  For large simulation cells, a good choice is to select 
a trial step size yielding a local acceptance ratio around 50\%, and to
maximize the number of intermediate (cheap) trial moves $M$ under the
constraint that the global acceptance ratio does not fall below 50\%.
The resulting $M$ depends on the accuracy of the energy estimates.

Depending on the system (i.e., metal or insulator, crystalline, disordered
solid, or liquid), there are different types of Monte Carlo moves
for which approximate energy differences can be computed efficiently
and accurately.  In metals, for example, a MC move localized in real
space can be combined with an embedded-atom potential, while a MC move
localized in reciprocal space is more appropriate in conjunction with
plane-waves {\it ab-initio} calculations.  Here, we present an approach
designed for semiconductors and insulators, and apply it to models of
amorphous silicon described by the tight-binding interaction developed
by Goodwin, Skinner and Pettifor \cite{gsp} and modified by Kwon 
{\it et al.} \cite{kwon}.

Within the tight-binding formalism, the total energy can be written as
\be
E = \sum_{i} \langle \Psi_i \vert H \vert \Psi_i \rangle + E_r +E_0
N_a
\ee
where the first term represents the electronic contribution of the
energy given by the sum of the occupied eigenvalues of the tight-binding
Hamiltonian $H$, the second term represents the ion-ion repulsion as
well as the correction from the double counting of the electron-electron
interaction and $E_0$ is a constant energy shift per atom, and $N_a$
is the total number of atoms in the system.  Details of the potential
and parameters used here can be found in Ref.~\cite{kwon}.

We want to compute the approximate energy differences between two
systems that are close to each other.  Because amorphous silicon is a
semiconductor, we keep the difference between the two configurations
localized in real space.  Consider two configurations A and B with
total energies $E_A$ and $E_B$ respectively, such that the atomic
coordinates differ only in a small region around a point $\vec{x}_m$;
for instance, in a Monte Carlo approach with single-atom displacements,
it is convenient to select $\vec{x}_m=(\vec{x}_i+\vec{x'}_i)/2$, the
geometrical center between the ``active'' region before and after
the displacement~\cite{whyM}.  A spherical localization volume is
constructed around $\vec{x}_m$, large enough to include the displaced
atoms or atoms and its local environment, following the {\em principle
of nearsightedness} of an equilibrium system~\cite{kohn} which asserts
that the effects on the deformation on atoms outside the localization
volume is small.  As discussed below, for the model system used here,
a sphere including about a hundred atoms around $\vec{x}_m$ is found to
be sufficient.

This discussion may be written formally.  Denoting I and II as the regions
inside and outside the localization volume, the exact energy difference
can be written as
\begin{eqnarray}
\delta E_{AB}&=& E_B-E_A \nonumber \\
&=& ({E_B}^I+{E_B}^{II})-({E_A}^I+{E_A}^{II}) \nonumber \\
&=& ({E_B}^I-{E_A}^{I})+({E_B}^{II}-{E_A}^{II}). \nonumber
\end{eqnarray}

To obtain an estimate of the approximate energy difference, we assume
that the contribution from region II for both configurations is be
the same. This is a good approximation as long as the radius of the
localized volume is greater than the localization length of the density
matrix of the system. The approximate energy difference $\delta \tilde{E}_{AB}$ 
is therefore simply given by $\delta \tilde{E}_{AB} = {E_B}^I-{E_A}^{I}$.
Furthermore, if the localization volume is big enough and remains
stationary during the atomic movement, the boundary effects too can
be neglected. We construct a local Hamiltonian for the localization
region I before and after the movement with open boundary condition
and calculate the band energies for the corresponding local Hamiltonian, the 
difference of which would give us the approximate change of the band 
energy in the two configurations.  Typically in our calculations we have
used volumes with a radius of 7.5 \AA, which roughly corresponds to a
hundred atoms.

We emphasize that although the quality of the approximation is strongly
related to the degree of localization, the algorithm works even if there
are states in the gap. In this case, the error in the estimate increases
but will be corrected for appropriately in the final MC step using the
full energy calculation. In this sense, the algorithm is much more stable
under the presence of defects than a typical order-N MD.

We test this scheme on three sample configurations of {\it a}-Si with 300,
500 and 1000 atoms, respectively.  The initial cells were prepared using
the optimized Wooten-Winer-Weaire\cite{www} bond-switching algorithm
described in Ref.~\onlinecite{highQ}. This method, which uses a Keating
interaction potential~\cite{keating}, produces the best {\it a}-Si models
to date, from a structural and electronic point of view.  

\begin{figure}
\epsfxsize=8cm
\epsfbox{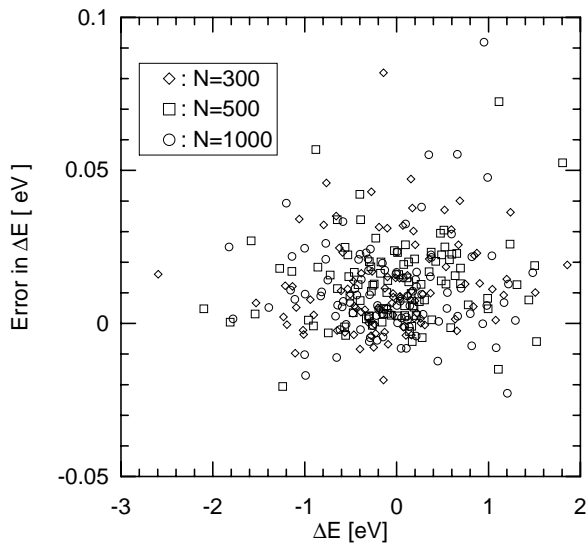}

\vspace{0.2cm}
\caption{
Accuracy of the O(1) estimate of the energy difference between two similar
configurations as described in the text.  The exact energy difference 
$\delta E$ between two configurations, calculated using a direct 
diagonalization method, is plotted horizontally.  The difference between 
the exact and estimated energy differences is plotted vertically.  The 
different symbols stand for the system sizes indicated in the figure.}
\label{o1_ener}
\end{figure}

First, we establish the validity of our approximate calculation of
the energy difference between two nearby configurations.  As shown
in Fig.~\ref{o1_ener}, where we compare the approximate energy,
$\delta \tilde E_{AB}$, with the exact energy difference obtained by
direct diagonalization of the full cell, the error implicit in this
approximation is about 0.03 eV, sufficiently small for the following
simulations.  The exact value of the error depends, of course, on a
number of variables: the distance between the original and the modified
configuration, the size of the localization volume, and the local strain
(which might delocalize some electronic states).  The error appears,
however, independent of the size of the simulation cell, which is to
be expected because of the aforementioned principle of nearsightedness.
For the results plotted in Fig.~\ref{o1_ener}, we selected a Monte Carlo
step of 0.4~\AA, much larger than the value we used in our Monte Carlo
simulations.

The efficiency of the various methods is determined by how fast each
samples the phase space. One way to measure this, is to estimate the
thermalization time from a configuration which is out of equilibrium.
Starting from the three configurations described above, which are
optimized for the Keating potential at zero K, we relax the atomic
positions at 300 K using two MC algorithms and an MD run.  In the first
MC approach, dubbed the standard MC approach, a trial configuration
$\vec{X}'_{i+1}$ is obtained from the old configuration $\vec{X}_i$ by
adding a gaussian random number to each of the $3N$ degrees of freedom.
The change in total energy $\delta E=E(\vec{X}'_{i+1})-E(\vec{X}_i)$ is
calculated, requiring a full total energy computation.  Next, the trial
configuration is either accepted or rejected following a Metropolis
probability as described in Eq. (\ref{eq:metrop}).  The spread of the
gaussian distribution is chosen so to maximize the diffusion in phase
space.  For our simulations with $N=300$ atoms, the atomic step size is
0.01 \AA; it decreases with increasing system size to about 0.0025 \AA\,
for a 1000-atom simulation.

The second method approach, called the multi-local approach, is the Monte Carlo
scheme proposed here: multiple local displacements, accepted or rejected
based on an estimate of the energy difference, followed by a single full
total energy calculation to recover correct sampling of the Boltzmann
distribution. Here, the size $M$ of the sequence of local moves and the
size of the atomic displacements are chosen so to maximize the diffusion
in phase space; the values we used are $M=300$ and a step size of 0.1 \AA.

The standard MD simulation is performed at constant temperature, by
rescaling the velocities at every time step. This modifies the trajectory
but ensures a faster convergence than a more flexible constraint dynamics,
damping considerably the thermal oscillations.  In MD, the forces also
need to be computed, in addition to the total energy. This requires the
evaluation of the eigenvectors of the tight-binding Hamiltonian, a step
which is not necessary in MC, and which increases the computational
cost as well as the memory requirements significantly. In spite of these
additional costs, MD is much faster than standard MC, since it is a
second-order scheme which makes use of inertia.

\begin{figure}
\epsfxsize=8cm
\epsfbox{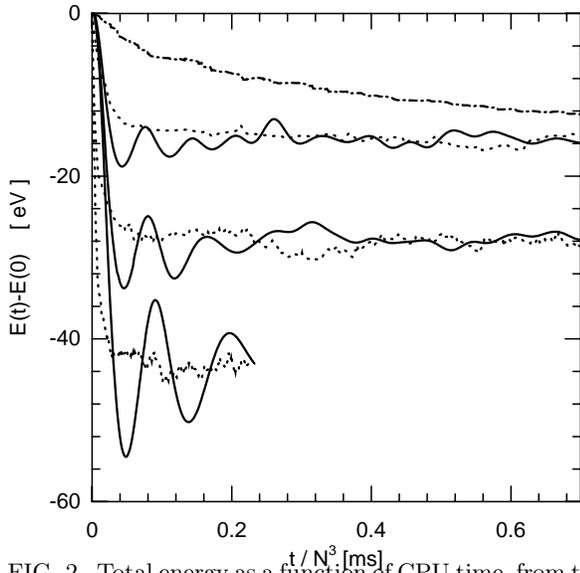}
\caption{Total energy as a function of CPU time, from top to bottom
for system sizes $N=300$, 500 and 1000. The dashed curves correspond
to results obtained with the non-local MC approach, the solid lines are
obtained with molecular dynamics. For the 300-atom model, the total
energy with the standard MC approach is plotted as the dash-dotted line.}
\label{fig:E}
\end{figure}

We compare the computational efficiency of these methods in figure
\ref{fig:E}, which shows the evolution of the total energy as a function
of CPU time on a fast workstation (DEC Alpha, 21264 architecture, 667 MHz)
for the methods applied to three different size cells. All curves
are fitted to the function
\begin{equation}
E(t)-E(0)=a\left( 1-exp(-t/\tau)\,cos(\omega t-\phi)\right) .
\end{equation}
Here, the initial excess energy is given by $a$, the thermalization
time by $\tau$, and the oscillatory behavior in MD by $\omega$ and
$\phi$; in the fits to the MC data, we took $\omega=\phi=0$ since MC
inherently does not show oscillatory behavior. Important for us are
the thermalization times. The fitted parameters for the multi-local MC
approach are $\tau/N^3=21$, 8.0 and 6.2 $\mu$s, respectively, for system
sizes $N$=300, 500 and 1000 atoms. For MD, we find $\tau/N^3=32$, 35 and
49 $\mu$s, respectively. (The increase in $\tau/N^3$ for MD is probably
due to the cache misses.)  Clearly, the multi-local MC outperforms MD
with an increasing factor as a function of system size; it is close to
an order of magnitude faster for the 1000-atom cell.

\begin{figure}
\epsfxsize=8cm
\epsfbox{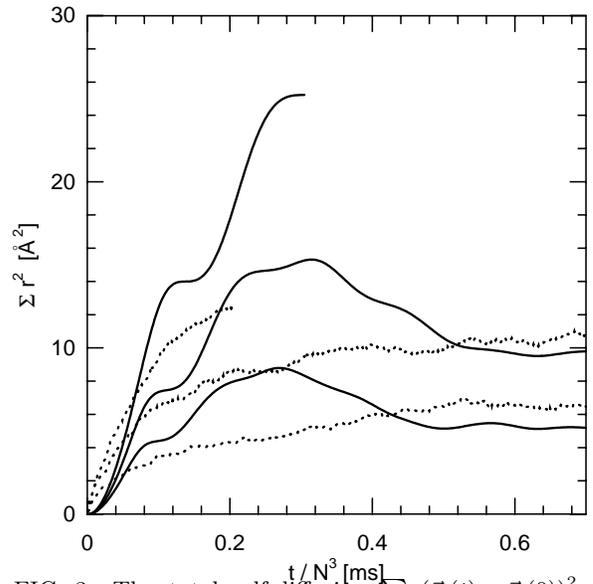}
\caption{The total self-diffusion
$\sum_i \left(\vec{r}_i(t)-\vec{r}_i(0)\right)^2$ as a function of CPU
time, for system sizes $N=300$ (bottom), $N=500$ (middle) and $N=1000$
(top).  The solid curves correspond to MD and the dashed lines to the
multi-local MC.
}
\label{fig:diffusion}
\end{figure}

The total self-diffusion $\sum_i \left[\vec{r}_i(t)-\vec{r}_i(0)\right]^2$
corresponding to these runs is reported in Fig. \ref{fig:diffusion}.
The MD curves show oscillatory behavior due to inertia, while the
overdamped MC dynamics leads to smooth non-decreasing diffusion.
The sudden jumps at irregular intervals are caused by activated events.
The difference in the displacement at the end of the simulations tells
that MC and MD have not reached the same local basin.

The resulting atomic configurations show excellent structural and
electronic properties; for the largest sample, the angular spread is
10.6 degrees (with a nearest-neighbor cut-off anywhere between 2.7 \AA{}
and 2.9 \AA), and the band gap is clean. A detailed study of the
structural and electronic properties is in progress~\cite{pb}.

Besides its higher efficiency, the MC approach is also more flexible. It
can easily concentrate the computational effort on the activate parts
of a heterogeneous system, for instance, an interfacial area in a
heterostructure, or the hydrogen atoms in {\it a}-Si:H.  Furthermore,
the efficiency can be raised by allowing well-chosen complex moves that
involve more than single-atom displacements. In the case of {\it a}-Si,
for example, we could include bond transpositions that dominate the
low-temperature dynamics~\cite{barkema97}.

In summary, we have presented here a multi-local Monte Carlo procedure
that exploits the fact that good estimations of energy differences between
relatively close configurations can often be obtained cheaply. This
permits us to construct a two-stage algorithm which first accepts-rejects
a large number of moves based on this approximate calculation of the
energy and then corrects the accumulated errors with a final accept-reject
step using one full energy computation.  This method is highly flexible
and can easily be adapted to a wide-range of problems including those
for which order-N molecular dynamics is inapplicable.

We have also presented an implementation of this algorithm for
thermalization of  tight-binding {\it a}-Si models. In this case, the
multi-local MC outperforms MD in the exploration of phase space by up
to an order of magnitude.

NM acknowledges partial support from the National Science Foundation
under grant number DMR-9805848. PB acknowledges support by NWO within
the priority program ``Solar cells in the 21$^{st}$ century''.


\begin{thebibliography}{*99}
\bibitem{whyM}
One aspect that is important for the proof of correctness of the Monte
Carlo approach is that the localization region is centered around
$\vec{x}_m$, thus independent whether the starting point was $A$ or $B$,
i.e., $\delta{\tilde E_{AB}} = -\delta{\tilde E_{BA}}$. If for instance
the localization region would have been centered around $A$, this property
would not hold any longer, and detailed balance would be violated.
\end{thebibliography}
\end{document}